\documentclass[aps,amsmath,amssymb,twocolumn,showpacs]{revtex4}
\usepackage{ulem}
\usepackage{color}

\usepackage{verbatim}
\usepackage{graphicx}
\usepackage{bm}
\def\pmb#1{\setbox0=\hbox{#1}
\kern-.025em\copy0\kern-\wd0 \kern-.05em\copy0\kern-\wd0
\kern-.025em\raise.0433em\box0}

\newcommand{\beq}{\begin{equation}}
\newcommand{\eeq}{\end{equation}}
\newcommand{\ba}{\begin{eqnarray}}
\newcommand{\ea}{\end{eqnarray}}

\input epsf

\newif\iffigures

\figurestrue 

\begin{document}

\title[]{Neolithic stone settlements as locally resonant metasurfaces}

\author{S. Br\^ul\'e$^{1}$, B. Ungureanu$^{2}$,
S. Enoch$^{1}$, S. Guenneau$^{3}$}
\affiliation{$^1$ Aix$-$Marseille Univ, CNRS, Centrale Marseille, Institut Fresnel, Marseille, France}
\affiliation{$^2$ LAUM, CNRS UMR 6613, Le Mans Université, 72085 Le Mans, France}
\affiliation{$^3$ UMI 2004 Abraham de Moivre-CNRS, Imperial College London, SW7 2AZ, UK} 
\begin{abstract}

We study the dynamic surface response of neolithic stone settlements obtained with seismic ambient noise techniques near the city of Carnac in French Brittany. Surprisingly, we find that menhirs (\textcolor{black}{neolithic human size} standing alone granite stones) with an aspect ratio between 1 and 2 periodically arranged atop a thin layer of sandy soil laid on a granite bedrock exhibit fundamental resonances in the range of 10 to \textcolor{black}{25} Hz. We propose an analogic Kelvin-Voigt viscoelastic model that explains the origin of such low frequency resonances. We further \textcolor{black}{explore} low frequency filtering effect with full wave finite element simulations. Our numerical results confirm the bending nature of fundamental resonances of the menhirs and further suggest additional resonances of rotational and longitudinal nature in the frequency range 25 to 50 Hz.
Our study \textcolor{black}{thus} paves the way for large scale seismic metasurfaces consisting of granite stones periodically arranged atop a thin layer of regolith over a bedrock, for ground vibration mitigation in earthquake engineering.  


\end{abstract}
\maketitle

Physicists study
by means of sources generating waves,
the dynamic response of so-called metasurfaces i.e. plate-made structures with pillars, with different geometrical characteristics and with different chemical compositions in order to obtain unusual physical phenomena. The \textcolor{black}{nature of the} source signal can be electromagnetic, acoustic, elastic waves,
etc. The plates (in a broad sense, thus encompassing substrates) can be nanometric, micrometric, centimetric, decametric, plurimetric objects \cite{lalanne1999design,maier2005plasmonics,zhu2017traditional,achaoui2011experimental,liu2012wave,pourabolghasem2014physics,colombi2016forests,palermo2018hybridization,brule2019role,ungureanu2019influence,CRMECA_2022__350_G2_237_0}.

In this Letter, we are interested in hectometric objects interacting with seismic waves and more precisely, seismic ambient noise \cite{sesame2005HVSR}. We illustrate the dynamic response of a granite site, the individual response of a menhir (\textcolor{black}{a prehistoric human size} standing alone granite stone) and the overall dynamic response of the stone grid. The purpose of this study is simply to test on field a real full-scale device made up of an exceptional geometric arrangement of stones.

Here we have recorded the seismic ambient noise by means of a 3D sensor. The signal processing consists of an HVSR method (Horizontal to Vertical Signal Ratio) and a spectral analysis of the three components of the sensor. By comparing the signals recorded at specific locations (on top of a menhir, between two rows of menhirs, etc.), the modes of interest are selectively  identified. These modes are linked to ground vibration, local resonance of a single menhir, global response of the lattice of menhirs, etc. We consider that the bending mode is dominant.

\begin{figure}
\centering
\includegraphics[width=9cm]{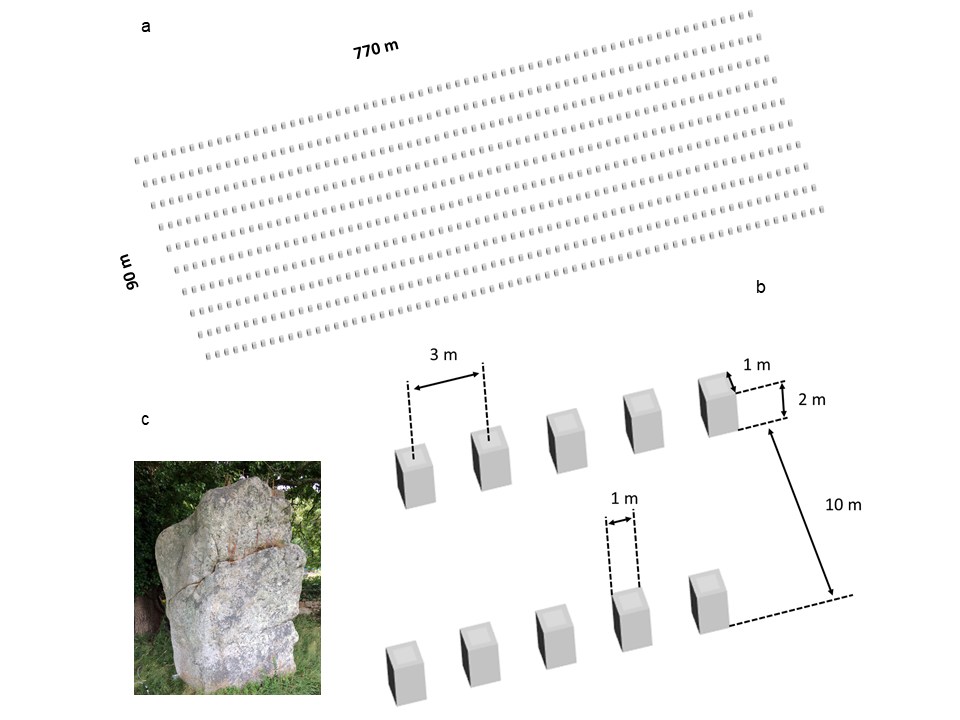}
\caption{Overall plate-model with alignments inspired from Carnac Stones in Britany. (b) Zoom view on elementary stone pillars. (c) Photograph of a typical menhir on the site.}
\label{figapl1}
\end{figure}

We have also defined a simplistic geometric plate model with square section pillars in order to compare the theoretical band diagrams \textcolor{black}{(computed with the commercial finite element package Comsol Multiphysics)} with the on-site measurement of the overall dynamic response of the 2D cubic lattice (Fig. \ref{figapl1}). To adjust the theoretical frequencies with the frequencies measured on field, we \textcolor{black}{implement} a classical soil-structure interaction (SSI) approach usually invoked for the rigid foundations of buildings. The  principles of SSI have been defined almost 50 years ago by Merrit and Housner \cite{merritt1954effect,housner1957interaction}. These seminal papers show the quantitative effect that foundation compliance has on the maximum base shear force and the fundamental period of vibration in typical tall buildings subjected to strong motion earthquakes.


The study takes place on the neolithic stone settlement of Carnac in Western part of France.
Visionary research in the late 1980's based on the interaction of big cities with seismic signals and more recent studies on seismic metamaterials, made of holes or vertical inclusions in the soil, has generated interest in exploring the multiple interaction effects of seismic waves in the ground and the local resonances of both buried pillars and buildings \cite{brule2019emergence,ungureanu2019influence,achaoui2017clamped,colombi2016forests,colombi2016seismic,colombi2017elastic,achaoui2016seismic,krodel2015wide,miniaci2016large,zeng2020subwavelength}.
Interestingly, the idea of a dense urban habitat with high-rise buildings has already \textcolor{black}{been proposed} in the past as perhaps in the medieval city of Bologna in Italy with its tall towers \cite{brule2019role}.
More recently, authors studied the concept of `seismic shield' properties of foundations of podium of roman-italic temples in central Italy with the case study of temple B in Pietrabbondante \cite{diosono2022seismic}.

The Carnac stones are an exceptionally dense collection of megalithic sites in Brittany in northwestern France, consisting of stone alignments (rows), dolmens (stone tombs), tumuli (burial mounds) and single menhirs (standing alone stones). More than $3000$ prehistoric standing stones were hewn from local granite and erected by the pre-Celtic people of Brittany.
Most of the stones can be found within the Breton village of Carnac. The stones were erected at some stage during the Neolithic period, probably around $3300$ BCE, but some may date to as early as $4500$ BCE.
The overall model is described on the basis of the two famous alignments of M\'enec and Kermario. For instance, the Kermario alignment is made up of approximately $980$ menhirs distributed in 10 rows, over a distance of approximately $1100$ m over approximately $100$ m wide. This set is generally oriented along a southwest-northeast

To set-up the model, we arbitrarily selected $10$ rows with a spacing of $3$ m between each stone. The height of the stones is $H_M=2$ m and their composition is local granite (density = $2.65$ $10^{3}$ kg.m$^{-3}$) as well for the plate-model. The basic model is made of a set of \textcolor{black}{stress-free cylindrical elements with a square-section ($B\times B=1\times  1$ m), which are clamped at one end} (see Fig. \ref{figapl1}).

\begin{figure}
\centering
\includegraphics[width=9cm]{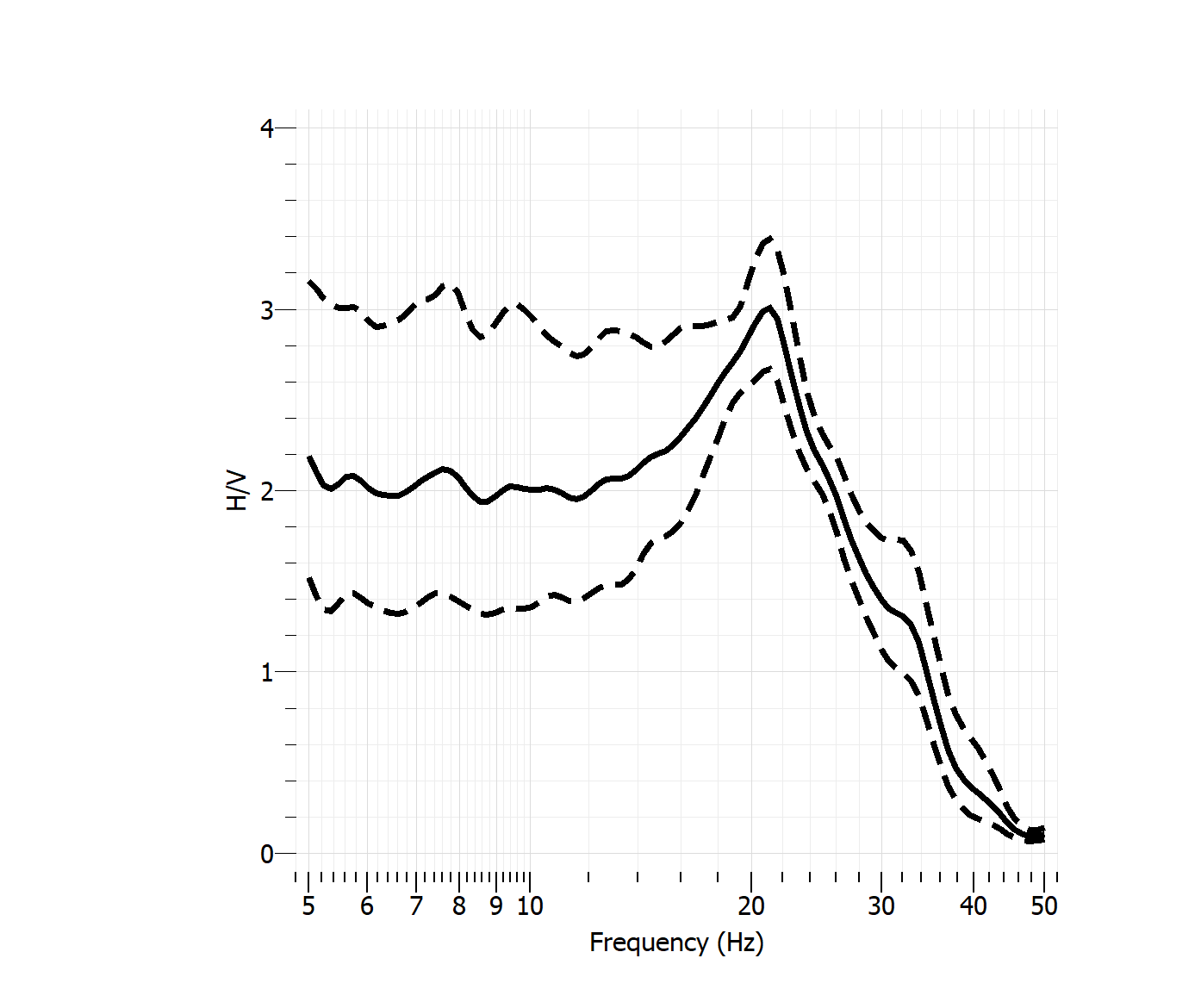}
\caption{HVSR spectral ratio versus frequency. The measurement was held between two rows of menhirs. The solid lines represent the average H/V curves, and the dotted lines show the ranges of the standard deviations. The sensor is located halfway between two standalone stones. \textcolor{black}{The resonant frequency is $20.6$ Hz $\pm 1.2$ Hz and it is attributed here to the top soil lying on the granite.}}
\label{figapl2}
\end{figure}

\begin{figure}
\centering
\includegraphics[width=8cm]{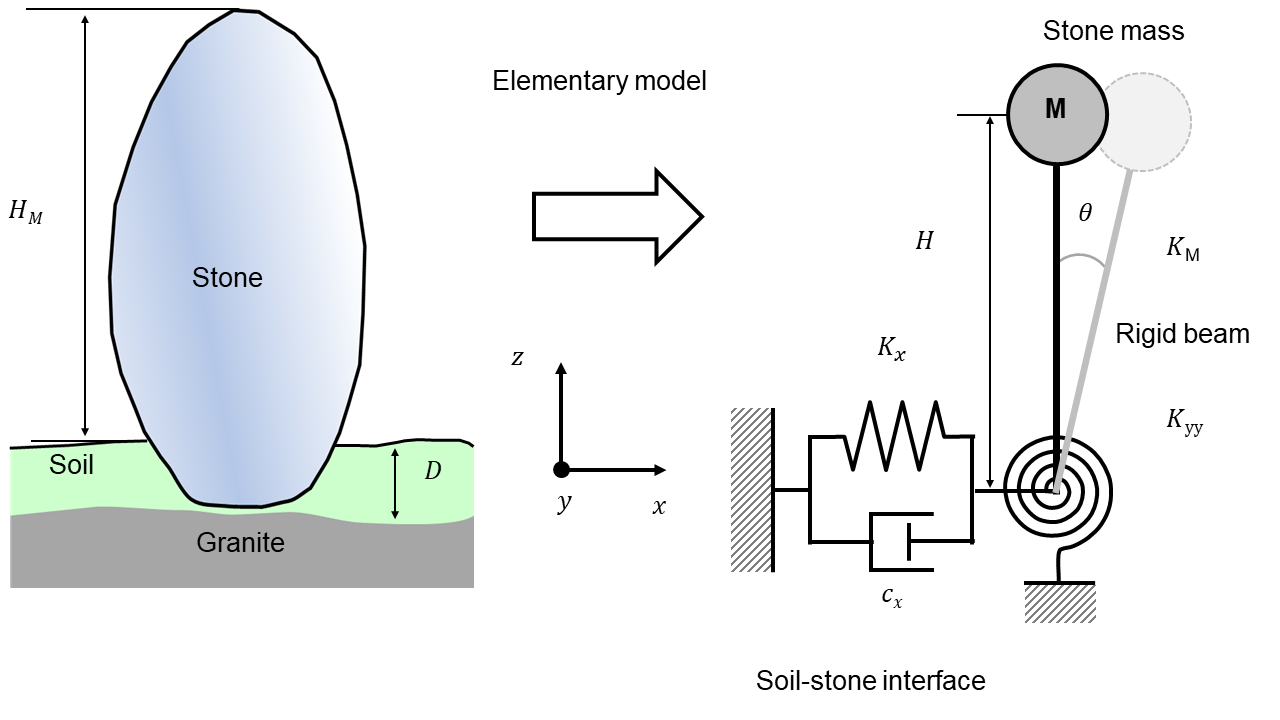}
\caption{The 1D locally resonant model. $D$ is the topsoil thickness. $H_M$ and $H$ are respectively the height of the menhir and the model ($H=2/3 (H_M)$). Soil-stone interface is modelled with an analogic Kelvin-Voigt viscoelastic model ($K_x$, horizontal stiffness, $c_x$, viscosity of the damper). $M$ is the mass of the stone, $K_M$ is the equivalent stiffness for the bending mode and $\theta$ is the rotation around $y$-axis, linked to $K_{yy}$, the rotational stiffness of the model.}
\label{figapl3}
\end{figure}

The local site conditions lead us to detail the initial clamped-free model. Indeed, standing stones are mainly placed on a granite base with a few tens of centimeters of topsoil (Figure \ref{figapl3}). We define a topsoil thickness of $0.5$ m, at least, for parameter D in Figure \ref{figapl3}. This depth is confirmed by archaeological studies \cite{hinguant2017menhirs} but also by the HVSR measurements carried out on site. In fact, the peak frequency for the soil is identified around 20.6 Hz (Fig. \ref{figapl2}) between two rows of stones. Assuming a shear wave velocity $V_s$ of $150$ m/s for the topsoil, we obtain $H_{soil}\approx 1.8$ m. D is assimilated to $H_{soil}$. The formula for the fundamental frequency $f_{0_{soil}}$ of a 1D soil model laying on a seismic substratum is:
\begin{equation}
    f_{0_{soil}}=V_s/(4.H_{soil})
\end{equation}

1D ground response analysis is based on the assumption that all boundaries are horizontal and that the response of a soil deposit is predominantly caused by SH-waves propagating vertically \cite{brule2014experiments}.


\begin{figure}
\centering
\includegraphics[width=8cm]{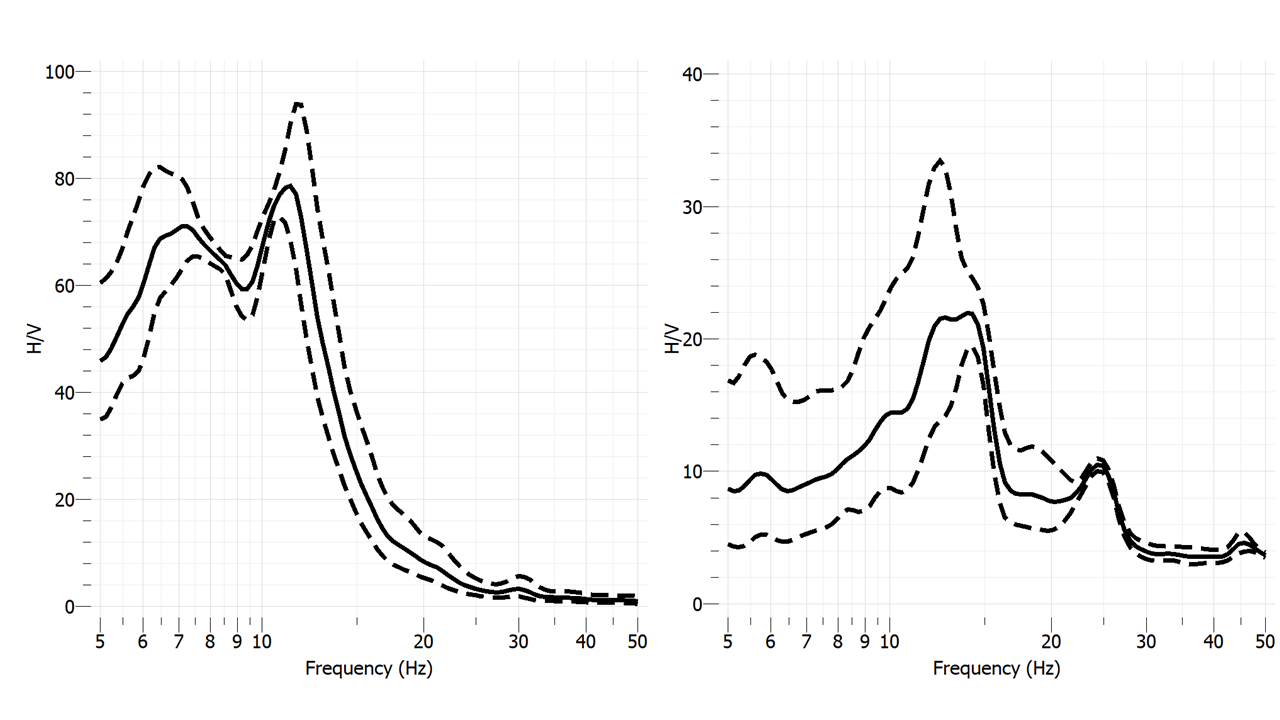}
\caption{HVSR spectral ratio versus frequency. The solid lines represent the average H/V curves, and the dotted lines show the ranges of the standard deviations. The sensor is located at the top of the menhir. On the left, the stone is 2 m high and 1 m on the right. In the left panel, the first dominant frequency is $11.3 \pm 0.6$ Hz.  The \textcolor{black}{first dominant frequency on left-hand side can be assigned to the bending mode in the perpendicular direction, according to Fig. \ref{figaplBU1new}. A third resonance is seen at $29.8 \pm 0.6$ Hz that is close to that of the rotational mode in Fig. \ref{figaplBU1new}. }\textcolor{black}{In the right panel, the first dominant frequency is $12.9 \pm 1.7$ Hz. A third resonant frequency at $24.5 \pm 0.35$ Hz can be assigned to the bending mode in the perpendicular direction, according to Fig. \ref{figaplBU2new}. A fourth resonance at $45.6 \pm 1$ Hz is close to that of the rotational and longitudinal resonances in Fig. \ref{figaplBU2new}.} 
}
\label{figapl4}
\end{figure}

First, let consider a single degree-of-freedom elastic structure resting on a fixed base \cite{clough1975dynamics} with $K_M$ the bending stiffness for a clamped-free beam, and mass $M$ (figure 3). From structural dynamics, the undamped natural vibration frequency, $\omega$ (rad/s), and period, $T_M$ (s), of the structure are given by:
\begin{equation}
    T_M=2\pi/\omega=2\pi\sqrt{M/K_M}
    \label{eq2}
\end{equation}
and
\begin{equation}
    K_M=(3E_M I_M)/H^3  
    \label{eq3}
\end{equation}
Considering $H=2/3 H_M$ [6], i.e. $H=1.33$ m, $E_M=8 000$ MPa, $I=B^4/12=0.083$ m$^4$, we find $T_M=0.016$ s or $f_M=63.5$ Hz. The clamping condition for this menhir induces a non realistic high frequency for the fundamental bending mode. Let us compare this analytical result with field measures by means of HVSR (figure \ref{figapl4}).

Initiated by Nakamura in 1989 \cite{nakamura89HVSR} and widely studied to explain its strengths and limitations \cite{brule2013methode,bard2008site,bonnefoy2006h,pilz2009comparison}  the seismological microtremor horizontal-to-vertical spectral ratio (HVSR) method offers a reliable estimation of the soil resonant frequencies from the spectral ratio between horizontal and vertical motions of microtremors (figure 2). This technique usually reveals the site dominant frequency $f_{0_{soil}}$ but the interpretation of the amplitude of the HVSR is still an open question. The HVSR method was tested in geotechnical engineering too by virtue of its ease of use \cite{brule2013methode}. In restrictive conditions, as for homogeneous soil layer model with a sharp acoustic impedance contrast with the seismic substratum, this technique could be employed to estimate the ground fundamental frequency $f_{0_{soil}}$. Ambient vibration techniques are also used for the determination of dynamic parameters of buildings \cite{chatelain2013reliable}. Here, the method is employed to estimate both $f_{(0 soil)}$ of the tested site, the dominant frequency of the bending mode of the menhir and the frequencies associated to the global dynamic response of the set of stones. Seismic noise was recorded with a sampling rate of $512$ Hz for 10 min at each site, this is to ensure that there is adequate statistical sampling in the range (0.1–50 Hz), the frequency range of engineering interest (TAB. 1 in Supplemental Material). HVSR curves are obtained with the open source Geopsy Software. Low-noise 20-s signal windows are selected through a short-term average (STA)/long-term average (LTA) antitrigger with 2-s STA, 30-s LTA, and low- and high-threshold of 0.2 and 3. A 5 percent cosine taper is applied to both ends of each time window on each component. The spectrum of each component of each individual window is smoothed according to the Konno and Ohmachi smoothing method \cite{ohmachi1998filter}, using a constant of 40. Then, HVSR in each window is computed by merging the horizontal (north– south and east–west) components with a quadratic mean. Finally, HVSR is averaged over all selected windows.

Basically, the ratio of the H/V Fourier amplitude spectra is expressed with
\begin{equation}
    HVSR(\omega)=|F_H (\omega)|/|F_V (\omega)| \; ,
    \label{eq4}
\end{equation}
where $F_V (\omega)$ and $F_H (\omega)$  denote the vertical and horizontal Fourier amplitude spectra respectively.

Different types of measures have been carried out on field. The sensors were placed on two stones to measure their local resonance (figure \ref{figapl4}), at the foot of these to capture the signal emitted by themselves and at ground level, far from other stones to identify the specific geological site response (figure \ref{figapl1}).

The measured frequency $11.3 \pm 0.6$ Hz for the 2 meter high stone and $12.9 \pm 1.7$ Hz for the 1 meter high one. These values are much lower than the theoretical ones for a perfectly clamped elastic model beam (equations \ref{eq2} and \ref{eq3}). This difference can be explained by flexibility at the soil-structure interface, i.e. horizontal relative displacement and rigid body rotation (figure \ref{figapl3}). Thus, we propose to introduce soil-structure interaction (SSI) to specify the parameters of the elementary model. This model is derived in Supplemental Material.

\begin{figure}
\centering
\includegraphics[width=7cm]{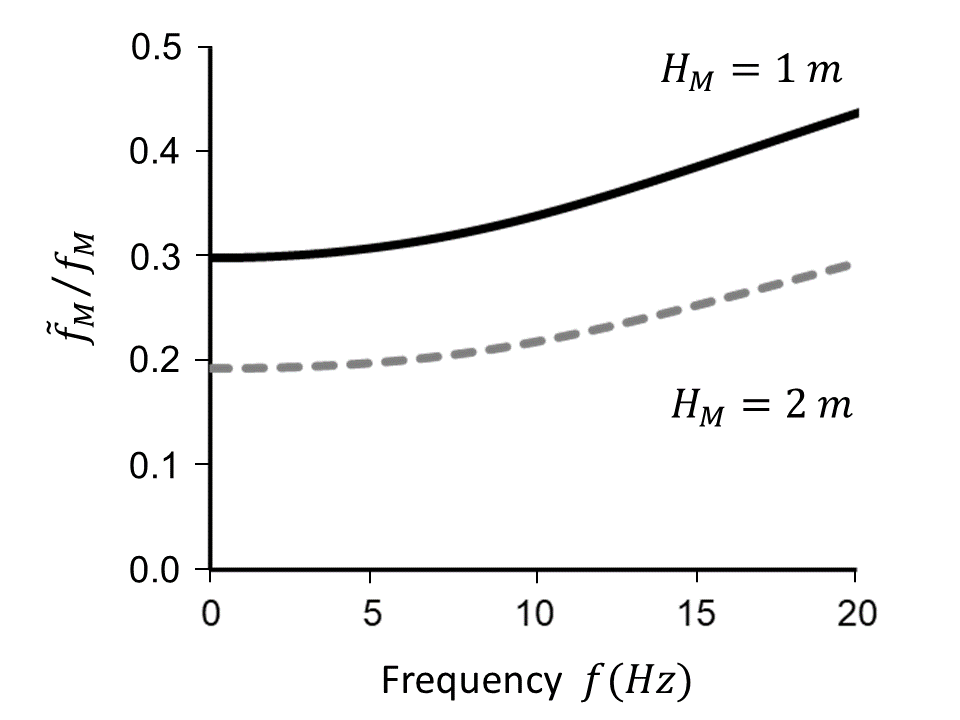}
\caption{Correction ratio $\tilde{f}_M/f_M$  for $B=1$ m, $H_M=1$ and $2$ m and $V_s=150$ m/s.}
\label{figapl6}
\end{figure}

We had previously obtained $f_M=63.5$ Hz. According to figure \ref{figapl6}, the SSI effect gives $\tilde{f}_M=63.5\times 0.19=11.97$ Hz. Thus the effect of the SSI is the lengthening of the period of the system. The equivalent height of the model $H_{eq}$ is given by
\begin{equation}
H_{eq}=\sqrt[3]{(3 E_M I_M (\tilde{T}_M )^2)/(4M\pi^2 )}
\label{eq9}
\end{equation}
We find $H_{eq}=4.06$ m (to be compared with $H=0.886$ m) and $H_{(M eq)}=6.08$ m (to be compared with $H_M=1.33$ m).

\begin{figure}
\centering
\includegraphics[width=8cm]{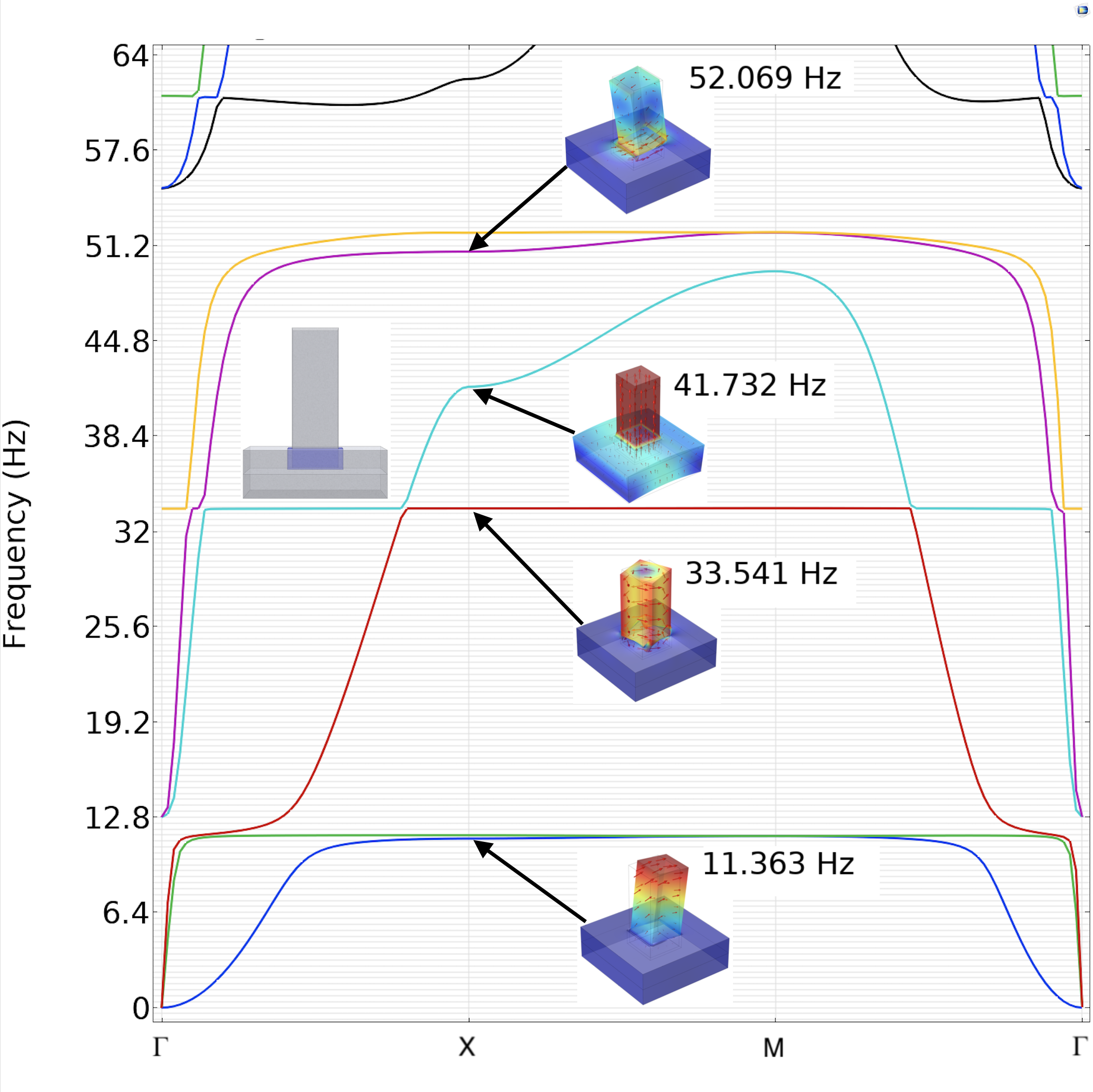}
\caption{Floquet-Bloch band diagram. The Young modulus, Poisson ratio and density of granite \& menhir are $E=2.5$ $10^4$ MPa, $\nu=0.2$, $\rho=2.3$ $10^3$ kg/m$^3$ and for soil $E=1.53$ $10^2$ MPa, $\nu=0.3$, $\rho=1.8$ $10^3$ kg/m$^3$. Menhir have height $2$ m above soil and depth $0.5$ m in soil, with square cross section $1$ m $\times$ $1$ m and centre to centre spacing of $3$ m. The plate consists of $0.5$ m of soil atop $1$ m of granite.
Importantly, part of mehnir in soil is surrounded by $0.05$ m of rubber ($E=2$ MPa, $\nu=0.49$, $\rho=1150$ kg/m$^3$). \textcolor{black}{We note the resonant frequency for the bending mode at $11.363$ Hz corresponds to that of the first dominant frequency in Fig. \ref{figapl3} (left). The resonant frequency of the rotational mode ($33.541$ Hz) is close to the third resonant frequency in Fig. \ref{figapl3} (left) whereas the longitudinal mode ($41.73$ Hz) is not detected in Fig. \ref{figapl3} (left). We note a complete stop band from $52$ Hz to $55$ Hz.}}
\label{figaplBU1new}
\end{figure}

\begin{figure}
\centering
\includegraphics[width=8cm]{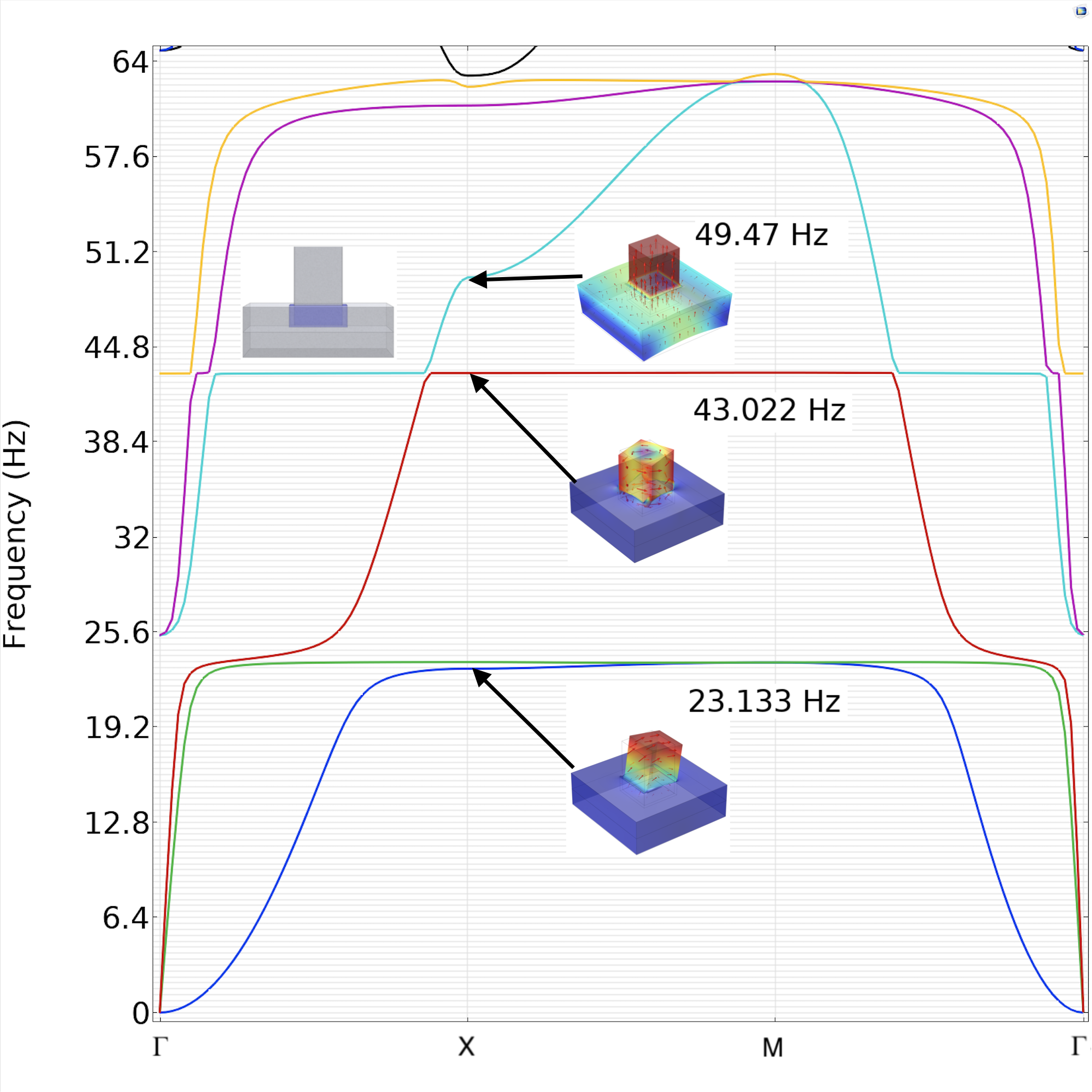}
\caption{Same parameters as in Fig. \ref{figaplBU1new}, except for stone's height which is $1$ m. \textcolor{black}{We note the resonant frequency for the bending mode at $22.133$ Hz is close to that of the second dominant frequency in Fig. \ref{figapl4} (right). 
The resonant frequencies of the rotational and longitudinal modes (at $43.022$ Hz and $49.47$ Hz, respectively) are close to the third resonant frequency in Fig. \ref{figapl3} (right).}} 
\label{figaplBU2new}
\end{figure}

\textcolor{black}{Regarding the numerical simulations, performed with the commercial finite element software Comsol Multiphysics, we note the presence of low-frequency resonances of a rotational nature in-between the fundamental bending mode and longitudinal mode resonances in Fig. \ref{figaplBU1new} and \ref{figaplBU2new}. The latter two are predicted by effective models such as in \cite{marigo2021effective}. However, the former has not been studied in great details yet, and seems to play a prominent role in the third resonance observed in Fig. \ref{figapl4}, as suggested by the band diagram in Fig. \ref{figaplBU1new}.}  

In conclusion, the last decade is marked by the emergence of metamaterials including seismic metamaterials. Well beyond the important scientific and technological advances in the field of structured soils, the concepts of superstructures dynamically active under seismic stress and interacting with the supporting soils (soil kinematic effect) and its neighbors is clearly highlighted.

\bibliography{menhir}
\end{document}